\documentclass[a4paper,10pt]{article}
\pdfoutput=1

\usepackage[english]{babel}
\usepackage[utf8x]{inputenc}

\usepackage{subfigure}

\usepackage{amsmath}
\usepackage{amssymb}
\usepackage{amsfonts}
\usepackage{wasysym}

\usepackage[ruled, linesnumbered]{algorithm2e}

\usepackage{ifpdf}
\ifpdf
\usepackage[pdftex]{color, graphicx}
\else
\usepackage{graphicx}
\fi

\newcommand\union{\cup}

\mathcode`|="326A          % | as a relation
\newcommand\auxfun[1]{\expandafter\newcommand\csname #1\endcsname{%
 \mathop{\hbox{\rm #1}}\nolimits}}
\newcommand\setsize[1]{\left| #1 \right|}

\auxfun{n}
\auxfun{nbrs}
\auxfun{msg}
\auxfun{trans}
\auxfun{est}
\title{\textit{Spectra}: Robust Estimation of Distribution Functions in Networks}

\author{Miguel Borges, Paulo Jesus, Carlos Baquero, Paulo Sérgio Almeida
\\\\\small{HASLab / INESC TEC \& Universidade do Minho}\\\small{Campus de Gualtar, 4710-057 Braga, Portugal}\\\texttt{\small\{mborges, pcoj, cbm, psa\}@di.uminho.pt}}
\date{Technical Report\\April, 2012}

\begin{document}

\maketitle

\begin{abstract}
	Distributed aggregation allows the derivation of a given global aggregate property from many individual local values in nodes of an interconnected network system.
Simple aggregates such as minima/maxima, counts, sums and averages have been thoroughly studied in the past and are important tools for distributed algorithms and network coordination. Nonetheless, this kind of aggregates may not be comprehensive enough to characterize biased data distributions or when in presence of outliers, making the case for richer estimates of the values on the network.

This work presents \textit{Spectra}, a distributed algorithm for the estimation of distribution functions over large scale networks. The estimate is available at all nodes and the technique depicts  important properties, namely:  robust when exposed to high levels of message loss, fast convergence speed and fine precision in the estimate. It can also dynamically cope with changes of the sampled local property, not requiring algorithm restarts, and is highly resilient to node churn.

The proposed approach is experimentally evaluated and contrasted to a competing state of the art distribution aggregation technique.
\end{abstract}

\section{Introduction}
	The ability to aggregate data is a fundamental feature in the design of
scalable information systems, which allows the estimation of relevant global
properties in a decentralized way in order to coordinate distributed
applications, or for monitoring purposes. Usual aggregates include
environment sensor data, such as temperature and humidity, and system
properties, such as load and available storage. 

Simple aggregates such as minima/maxima, counts, sums and averages have been
thoroughly studied in the past. Nonetheless, this kind of aggregates may not
be comprehensive enough to characterize biased distributions or in the
presence of outliers, making the case for  richer estimates of the values on
the network (e.g. probability density functions, histograms, cumulative
distributed functions), since statistical ordinary moments hide, in many
cases, changes in the property that are relevant to control decisions.

The amount of scientific work is relatively scarce in what concerns more
expressive aggregation metrics. A recent proposal within this
domain (\textit{Adam2})~\cite{adam2} claims to obtain estimates with a better precision than
in previous approaches. It is an algorithm for the estimation of discrete
cumulative distribution functions.

Despite the contribution, the proposal mentioned above is not fault tolerant
and is also not sensible to the continuous variation of the sampled
properties, for it demands the protocol to be restarted frequently in order
to achieve quasi-continuous monitoring. Besides, the approach does not admit
loss or duplication of messages.

Having this scenario as a starting point, this work presents
\textit{Spectra}, a distributed algorithm for the estimation of distribution
functions over large scale networks. Its core advantages are resilience to
message loss, high convergence speed and high precision of the estimate. It
also supports changes of the sampled property and churn. All this is
achieved without requiring the protocol to be restarted.

In detail, \textit{Spectra} enables the estimation of the cumulative
distribution function (CDF) of a given property at all nodes. This allows
nodes to take advantage of having a broader view of the property on the
network: they may exclude outliers or monitor particular quantiles of a
property. Also, each node of the network has a local vision of the global
state of the property, thus allowing them to make decisions based on local
knowledge. 

Since the approach used by \textit{Adam2} is the most resemblant to the proposed work, we have  included simulation results that support and validate our
approach along with a comparison with the \textit{Adam2}
algorithm.

In the next section we make a  short overview of the state of the art work
on the context of distribution aggregation. In Section 3 we briefly state
the system model used on this work. Section 4 presents the
\textit{Spectra} algorithm and after we show the evaluation results,
contrasting with \textit{Adam2}. Last section draws conclusions on the work
and presents a few perspectives about future research directions.

\section{Related Work}
	In the last decade, several distributed aggregation algorithms have been
proposed to estimate the value of scalar properties (e.g. network size).
Existing techniques can be divided in different classes, providing different
characteristics in terms of performance (time and message load) and
robustness, mainly as: hierarchy-based (or tree-based), averaging (or
gossip-based), sketches and sampling approaches. A wide and comprehensive
overview of the current state of the art on distributed aggregation
algorithms is provided in \cite{Jesus:2011vo}. %the following survey .

Hierarchy-based approaches
\cite{Madden:2002p2402,Motegi:2005p1017,Birk:2006p3907} rely on a
process to aggregate data along a pre-established hierarchical
routing structure (commonly a tree), producing the result at the root. This
kind of technique is usually applied to Wireless Sensor Networks (WSN) due
to its energy efficiency, despite being highly sensitive to failures. 
Some algorithms
\cite{Ganesh:2007p745,Cheng:2010p8605,Mane:2005p659} are found collecting
samples and applying an estimation method to obtain a rough approximation of
the size of a membership. This type of scheme is lightweight in terms of
message load, as only a partial number of nodes might be asked to
participate in the sampling process, but is also inaccurate and produce the
result at a single node. Moreover, several rounds might be required to
collect a single sample, especially when sampling is performed through
random walks like in~\cite{Ganesh:2007p745,Mane:2005p659}, thus being slow.
A more robust alternative is provided by algorithms that aggregate data
through multiple paths, such as those based on the use of sketches
\cite{Considine:2004p676,Nath:2004p1114,Fan:2010p9070,Baquero:vu}, enabling
all nodes to produce a result. These algorithms are fast (i.e. obtaining an
estimate in a number of rounds close to the diameter value of the network
graph), but not accurate. Another interesting alternative is provided by
\emph{averaging} techniques
\cite{Kempe:2003p700,Jelasity:2005p664,JenYeuChen:2006p3916,Jesus:2010ke,Almeida:vj},
which can reach an arbitrary accuracy, with the estimate at all nodes
converging to the correct result over time.

Most of the existing approaches allow the distributed computation of
aggregation functions, such as \textsc{count}, \textsc{average},
\textsc{sum}, \textsc{max/min}, and therefore the calculation of many scalar
values that can result from the combination of those functions. However,
they are unable to compute more complex aggregates which provide a richer
information about some property, such as the frequency distribution of an
attribute. In fact, few approaches are found in the literature that allow
the distributed estimation of statistical distributions
\cite{Greenwald:2004p7935,Shrivastava:2004p7854,Haridasan:2008p8141,adam2},
and those found exhibit robustness and accuracy issues. 

Algorithms like \cite{Shrivastava:2004p7854} and \cite{Greenwald:2004p7935}
require a tree routing structure to produce an approximation of the
distribution at the root, operating similarly to common tree-based
aggregation techniques. In particular, each node computes a quantile summary
(i.e. digest) holding the data from its sub-tree (e.g. range of values and
corresponding counts) which are built in a bottom-up fashion toward the
root. Like in classic tree-based approaches, a single failure may affect the
aggregation process, leading to the loss of the data from a subtree. 

A first gossip-based distribution estimation approach was proposed in
\cite{Haridasan:2008p8141}, randomly exchanging and merging finite lists of
bins (i.e. pairs with value and respective counter) between nodes.
Initially, the list of bins at each node is set with the initial input
value, and after several rounds all will produce an approximation of the
distribution of values (i.e. histogram). Different merging techniques were
considered by the authors, the one referred to as \emph{equi-depth} showed
to be the one with the best results (accuracy vs storage) compared to the
others. The \emph{equi-depth} method intends to minimize the counting
disparity between bins. In particular, upon reception, received and local
pairs are ordered and the pairs of consecutive values with the smallest
combined count are merged (i.e. counts are added and the new value results
from the weighted average) repeatedly until the desired number of bins is
obtained. This approach allows data to reach all nodes through multiple
paths (in this sense improving the robustness), but also gives rise to the
occurrence of duplicates that will bias the produced estimate. This problem
was acknowledged by the authors, arguing that it was better (i.e. simpler
and efficient) not to try to solve it.

Adam2~\cite{adam2} is a more recent gossip based approach to approximate
distributions, more precisely CDF. This approach is based on the application
of a classic averaging technique, namely Push-Pull
Gossiping~\cite{Jelasity:2005p664}, and at a high abstraction level it can
be simply described as the simultaneous execution of multiple instances of
this protocol. In more detail, Adam2 considers a fixed list of $k$ pairs
$(s_k,e_k)$, where $s_k$ is an interpolation point and $e_k$ is the fraction
of nodes with a value $x$ less or equal than $s_k$. Each node $i$ that
starts participating in the protocol initializes its list of pairs setting
$e_k = 1$ if $x_i\le s_k$ and $e_k = 0$ otherwise. Then, the Push-Pull
Gossiping process is applied, each node randomly picking a neighbor to
exchange their list of pairs and individually averaging the fractions
corresponding to each interpolation point. Over time, the fractions will (be
expected to) converge at all nodes to the correct value in each pair. Adam2
solves the duplication problem of the previous \emph{equi-depth} method,
considerably outperforming it according to the provided evaluation results.
Nevertheless, as will be showed in Section~\ref{sec:evaluation}, Adam2
inherits the ``mass loss'' problems of Push-Pull Gossiping, not converging
to the correct result even in fault-free scenarios~\cite{Jesus:2009p8115}. 

This work proposes a truly fault-tolerant and more accurate alternative,
with the fractions of each interpolation point effectively converging at all
nodes over time and simultaneously supporting dynamic changes. 

\section{System Model}
	\label{sec:model}

Our model assumes the existence of a large number of distributed processes
or nodes. Our goal is to estimate an accurate distribution of an attribute
over the network of processes with a robust aggregation strategy.

The assumptions stated bellow are defined for the purpose of evaluating the system.

The network of distributed nodes is modeled as a connected undirected graph
$\mathcal{G}(\mathcal{V},\mathcal{E})$, with the set $\mathcal{V}$
representing processes and the set $\mathcal{E}$ being bidirectional
communication links between processes. We represent the set of adjacent
nodes of node $i$ by $\mathcal{D}_i$.

The algorithm is executed synchronously as described in~\cite{nancy}(Chapter
2). Each node executes two procedures in lockstep each round: they begin
execution by generating messages to deliver to neighbors and sending them.
Afterwards, nodes compute their new state as a function of its current state,
the observed value and received messages from neighbors. Nodes do not have
global Ids and have only to distinguish the members of the set of neighbors.
This assumption, although not strict, is pertinent in a network with a massive number of sensors, where the number of computational units involved may turn intractable the assignment of unique Ids.

Message loss are taken into consideration and modeled as follows: per round, 
each sent message can be dropped according to a predefined uniform random probability.
In terms of dynamic changes (input values and churn), it is assumed that they occur 
at the beginning of each round (i.e. before the message generation procedure). 
Departing nodes are chosen uniformly at random, and it is assumed that they do not 
return to the network (i.e. leave forever). Arriving nodes connect to random points of 
the network (according to the considered topology) and establish a number of links 
matching the network properties (i.e. degree).

\section{Spectra -- Robust Distribution Estimation}
	In this section we describe a novel distributed algorithm, \textit{Spectra},
to estimate the distribution of a global attribute, more specifically its
Cumulative Distribution Function (CDF). A CDF can be approximated by a
mapping from points to the frequencies of the values that are less than
or equal to each point. More precisely, considering $n$ nodes and an input
value $x_i$ at each node $i$, the CDF of $x$ can be approximated by a set of
$k$ pairs $(s, e)$, where $s$ is an interpolation point and $e$ is the
fraction of values less or equal than $s$ (i.e., $e = \setsize{\{x_i | x_i
\le s\}}/n$).

Considering each pair $(s, e)$ in the CDF, it is possible to
estimate $e$ through an averaging algorithm as follows:
setting $1$ as the input value of each node $i$ that satisfies the condition
$x_i \le s$ and $0$ otherwise, the average of these input values will be
the fraction of nodes that fulfill the previous condition, i.e. it will be
$e$.  This means that $e$ can be computed as result of the execution of a
distributed averaging algorithm.

The main idea of the proposed algorithm is the combination of this
observation with the adaptation of a robust distributed averaging
algorithm, Flow Updating~\cite{Jesus:2010ke} (FU)\footnote{Flow Update is a distributed averaging algorithm that computes averages with the value each node observes. It exchanges estimates and flows with its neighbors. Flows are symmetric between two adjacent nodes and quantify the amount a recipient node should adjust its estimate in order to converge to the sender estimate. Through this technique, eventually all nodes converge to the same average.}, to work with vectors
instead of scalars, one component of the vector for each point of the CDF.

Simultaneously, whilst the algorithm is converging, a distributed
computation of the global minimum and maximum of the values is performed
to determine the interval in which the $k$ points of the CDF are estimated.

\auxfun{merge}
\auxfun{transform}

\begin{algorithm}[!t]
\AlFnt{\small\sf}
\SetKwBlock{Inputs}{inputs:}{}
\SetKwBlock{StateVars}{state variables:}{}
\SetKwBlock{StateTrans}{state-transition function:}{}
\SetKwBlock{MsgGen}{message-generation function:}{}
\SetKwBlock{EstFun}{estimation function:}{}
\SetKwBlock{IntFun}{interval interpolation:}{}
\SetKwBlock{MergeFun}{interval merging:}{}
\SetKwBlock{AdjFun}{vector transformation function:}{}
\SetAlgoVlined
\SetAlgoInsideSkip{bigskip}
%\SetAlCapSty{textnormal}
\DontPrintSemicolon
\Inputs{
	$x_i$, value to aggregate\;
	$\mathcal{D}_i$, set of neighbors\;
	$k$, number of interpolation points\;
}
\StateVars{
	flows: initially, $F_i = \{\}$ \tcc*[f]{mapping from neighbors to flow vectors}\;
	base frequency vector: initially, $\vec{v_i} = [1 | 0 \leq j < k]$\;
	interpolation interval: initially, $I_i = (x_i, x_i)$\;
}
\MsgGen{
	$\msg_i(F_i, \vec{v_i}, I_i, j) = (i, I_i, \vec{f}, \est(\vec{v_i}, F_i))$;
	\BlankLine	
	\textbf{with} $\vec{f} =
		\begin{cases}
    		F_i(j) & \textrm{if } (j,\_) \in F_i \\
    		\vec{0} & \textrm{otherwise}
  		\end{cases}$
}
\StateTrans{
	$\trans_i( F_i, \vec{v_i}, I_i, M_i ) = (F_i', \vec{v_i}', I_i')$
	\BlankLine	
	\textbf{with}
    \BlankLine	
	$I_i' = \merge(I_i \union \{ I | (\_, I, \_, \_) \in M_i \})$ \;
	$\vec{v_i'} = [\textrm{if } x_i \leq I_i'(j) \textrm{ then } 1
\textrm{ else } 0 | 0 \leq j < k]$ \;
	$F = \{j \mapsto -\transform(\vec{f}, I, I_i') | j \in \mathcal{D}_i
\land (j, I, \vec{f}, \_) \in M_i\} \union {}$\\
	$\hphantom{F={}} \{j \mapsto \transform(\vec{f}, I_i, I_i') | j \in \mathcal{D}_i \land (j, \_, \_, \_) \not\in M_i \land (j,\vec{f}) \in F_i\}$\;
	$E = \{ i \mapsto \est(\vec{v_i'}, F) \} \union {}$\\
	$\hphantom{E={}} \{ j \mapsto  \transform(\vec{e}, I, I_i') | j \in
\mathcal{D}_i \land (j, I, \_, \vec{e}) \in M_i\} \union {}$\\
	$\hphantom{E={}} \{ j \mapsto  \transform(\est(\vec{v_i}, F_i), I_i,
I_i')  | j \in \mathcal{D}_i \land (j, \_, \_, \_) \not\in M_i \} $\;
	$\vec{a} = (\sum\{\vec{e} | (\_, \vec{e}) \in E\}) / \setsize{E}$\;
	$F_i' = \{ j \mapsto \vec{f} + \vec{a} - E(j) | (j, \vec{f}) \in F \}$
}
\EstFun{
	$\est(\vec{v}, F) = \vec{v} - \sum\{\vec{f} | (\_, \vec{f}) \in F\}$
}
\MergeFun{
	$\merge(S) = (\min(\{l | (l, \_) \in S\}), \max(\{u | (\_, u) \in
S\}))$
}
\IntFun{
	$(l, u)(j) = l + j \times (u - l) / (k - 1)$
}
\AdjFun{
	$\transform(\vec{u}, I, I') =
	[\vec{u}(\max(\{0\} \union \{l | 0 \leq l < k \land I(l) \leq I'(j)\}))
		| 0 \leq j < k]$
}
\caption{Spectra: Algorithm to estimate CDF in distributed networks.}
\label{alg:fucdf_algorithm}
\end{algorithm}

This new algorithm is referred to as \textit{Spectra}, and its core is based
on the application of FU to estimate a CDF. The computation performed at
each node $i$ is detailed by Algorithm \ref{alg:fucdf_algorithm}.  The
algorithm adapts FU averaging to use vectors instead of scalars.  Namely,
the flows $F_i$ map for each neighbor a vector of flows (one for each point
in the CDF); $\vec{v_i}$ is a vector which contains the contribution of the
node according to the input value $x_i$, being used as the input to the FU
algorithm; and the estimation function yields a vector of estimates, the $k$
points of the CDF.

The algorithm does not assume knowledge of the global minimum and maximum
values. Instead, each node keeps a local knowledge of the minimum of maximum
known so far in the interpolation interval state variable ($I_i$). The
interval is sent in messages to neighbors, which merge the received intervals.
After $d$ rounds, where $d$ is the network diameter, each node $i$ will have
the the global minimum and maximum values in the $I_i$ variable.

The present algorithm computes an equi-width approximation of the CDF at $k$
equi-distant points in the interval between the global minimum and maximum
(although other variants are possible).  We use a notation where an interval
$I=(l,u)$ is indexed, i.e., $I(j)$, with $j$ from 0 to $k-1$, resulting in
the $k$ equi-distant points of interest from the lower to the upper value
(line 29).  In this paper we assume that the number of points, $k$, is fixed
and known to all nodes. It is possible to relax this assumption and derive a
system where $k$ can be adapted at execution time. 

Before global minimum and maximum convergence, the vectors calculated at
each node or in different rounds refer to different points. At each
iteration, as a new interval is computed by merging intervals in messages,
the algorithm needs to transform both the received vectors as well as the
vectors from the previous iteration, so that they are meaningful for the 
new (and potentially different) set of $k$ points. For that all vectors
involved in the execution of the algorithm are transformed from their old to
the new interval through the vector transformation function (line 31).
This function implements a simple heuristic to obtain the new vector, using
the value corresponding to the largest point not greater than the new
point (or the first in the vector if no such point exists).
Also, the vector of input values to FU, $\vec{v_i}$, is calculated at each
round according to the new interval (line 16).

At each round, in the message generation function (lines 9--11), a single type of message is sent, containing the self id $i$, its interpolation interval $I_i$, the flows vector $\vec{f}$ for each current neighbor $j$. Sent flows are set according to the current state and otherwise (initially or when a node is added) to $0$.

The state-transition function (lines 12--23) takes state (i.e., flows $F_i$, the node's base frequency vector $\vec{v_i}$, interpolation interval $I_i$) and the set of messages $M_i$ received by the node and returns a new state (i.e., flows $F_i'$, base frequency vector $\vec{v_i'}$ and interpolation interval $I_i'$). It computes the new interpolation interval setting the lower bound with the minimum of the received minima and does the upper bound with the maximum (line 27). Base frequency vector $\vec{v_i'}$ is computed from the new interpolation interval $I_i'$ and the initial value $x_i$. Then, the averaging steps are executed according to FU taking care to transform the involved vectors in order to apply the averaging process to matching interpolation points. These steps result in the creation of the new flows.

The self-adapting nature of the core averaging algorithm, Flow-Updating, on
\textit{Spectra} enables it to cope with the dynamic adjustment of all
involved vectors. In particular, \textit{Spectra} supports dynamic network
changes (i.e., nodes arriving/leaving), simply by adding/removing the flow
data associated to neighbors. Moreover,  it is also able to seamlessly adapt
to changes of the input value $x_i$ -- in this case simply by recomputing
the vector $\vec{v_i}$. This is sufficient to allow \textit{Spectra} to
operate in settings where the global maximum and minimum
do not change.

If the extreme values change due to dynamism, especially if
the maximum decreases and the minimum increases, the algorithm as it is will
not produce wrong results, but over an overly wide interval.
To tighten the interval to the interesting range between the new minimum and
maximum additional modifications must to be made.
This dynamic adjustment of the global extreme values (i.e., maximum and
minimum) is left for future work.

At each node $i$, the estimated CDF at the $k$ equi-distant points in the
interval $I_i$ is obtained by the estimation function (i.e.,
$\est(\vec{v_i}, F_i)$). Over time, the estimated frequency associated to
each point converges to the correct value. This is confirmed by the results
obtained from evaluation (see Section~\ref{sec:evaluation}). 

\section{Evaluation}
\label{sec:evaluation}
	The results presented in this work have been obtained using a custom made
simulator that implements the model defined in section~\ref{sec:model}.

We used two error metrics to quantify the fitness of the estimate to the underlying distribution. They are computed at every round. 

The basis of these metrics is the Kolmogorov--Smirnov statistic, that calculates the
maximum label-wise difference between the estimate and the distribution, as
presented in Equation~\ref{ksrn}. The metric is computed at every round $r$, for each node $n$. For every label $l$, the measure is given by the difference between the cumulative value of the real distribution and the cumulative value of the estimated distribution on node $n$ at round $r$.

\begin{align}
KS_r^n &= \max_{l \in Labels}|P(X \leq l)_r - \widehat{P(X \leq l)_r^n}| \label{ksrn} \\
KS_{max r} &= \max_{n \in N} (KS_r^n) \label{ksmax} \\
KS_{\mu r} &= \frac{1}{|N|} \sum_{n \in N} KS_r^n \label{ksmed}
\end{align}

We use global metrics for the whole network: one to reflect the worst node (see Equation~\ref{ksmax}) and another to reflect the average error (see Equation~\ref{ksmed}). Both equations are computed at every round $r$.

\subsection{Comparison with an existing approach}

To the best of our knowledge, this work is pioneer in the estimation of
statistical distributions with fault tolerant and robustness properties. 

In \textit{Adam2}~\cite{adam2}, whose protocol  is based on the
\textit{Push-Pull gossiping} averaging algorithm~\cite{pushpull}, presents a few
drawbacks stemming from the fact that it behaves poorly under message loss
and also because the simulator used on the above-cited work (PeerSim) does not emulate synchronous message exchange correctly, assuming atomic state changes \-- this behavior is, from
our point of view, unrealistic when considering real systems.

\textit{Adam2} partitions the range of the monitored property in a set of
interpolation points. Nodes start the algorithm with a pre-known minimum and
maximum and with equally spaced interpolation points. Probabilistically, new
instances are created every few rounds. These new instances re-compute those points based on the previous instance's points set, using
re-sampling heuristics that aim to minimize interpolation errors. In short, this operation
aims to concentrate interpolation points in the areas where frequency counts are
more prevalent.

In order to compare our approach with the strategy used on \textit{Adam2},
we assume that both minimum and maximum are previously known to all nodes
and the sampling points are evenly distributed between minimum and maximum.
These assumptions do not invalidate the usefulness of the re-sampling
heuristics presented, but help us compare the performance of both
approaches in a common frame of reference. Also, these heuristics are also applicable to \textit{Spectra} in a scenario with multiple instances, but that falls out of the scope of the present work.

We have simulated the following scenarios using a 1000 nodes random network with an average connectivity of 3, unless stated otherwise. The underlying initial
values follow a Normal distribution with mean 10 and variance 2. Results were averaged from 30 trials for each scenario.

\begin{figure}[t!]
\centering
\subfigure[Average KS error rate on a 1000 nodes random network comparing \textit{Spectra} and \textit{Adam2} convergence.]{
\includegraphics[width=.45\textwidth]{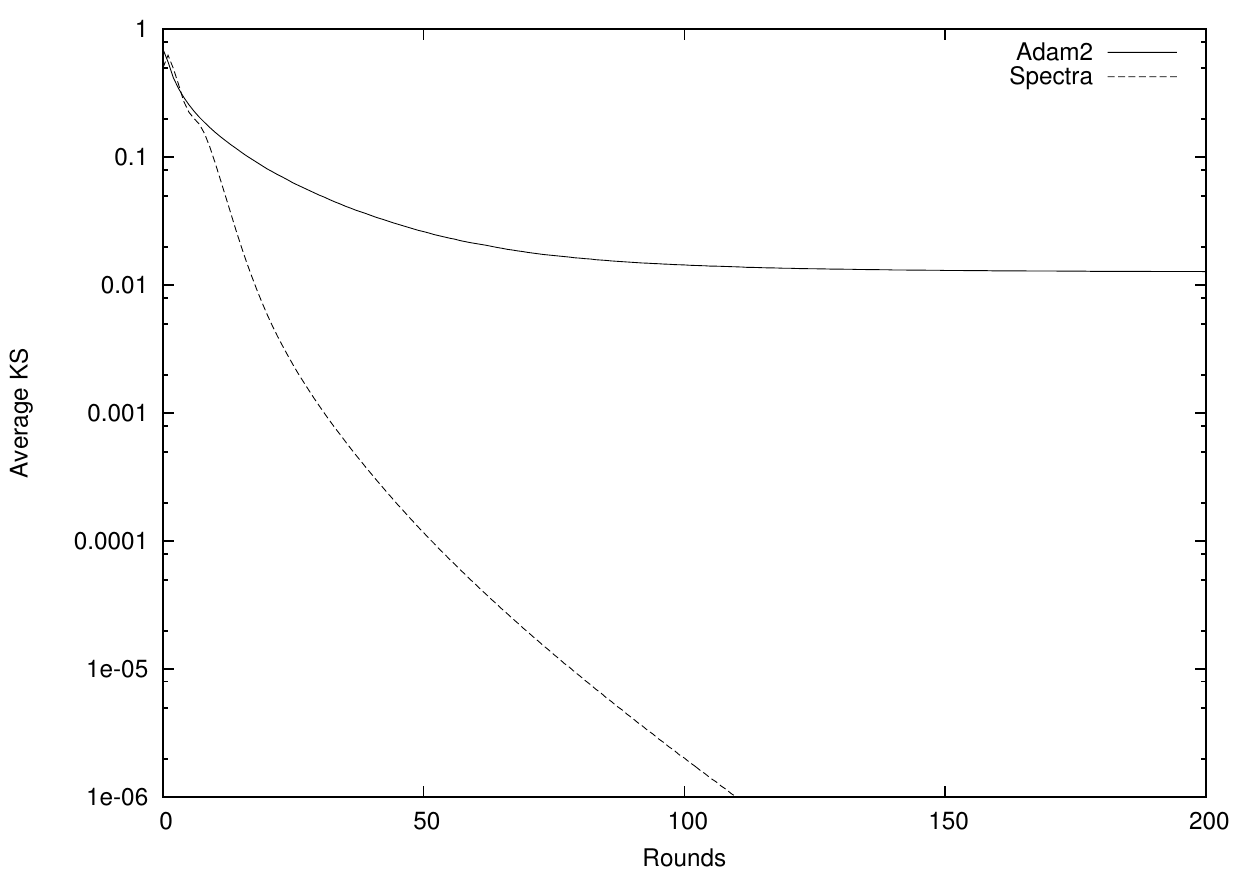}
\label{fig:adam2spectra}
}
\subfigure[Average KS error rate on a 1000 nodes random network comparing message loss effect on \textit{Spectra} and \textit{Adam2}.]{
\includegraphics[width=.45\textwidth]{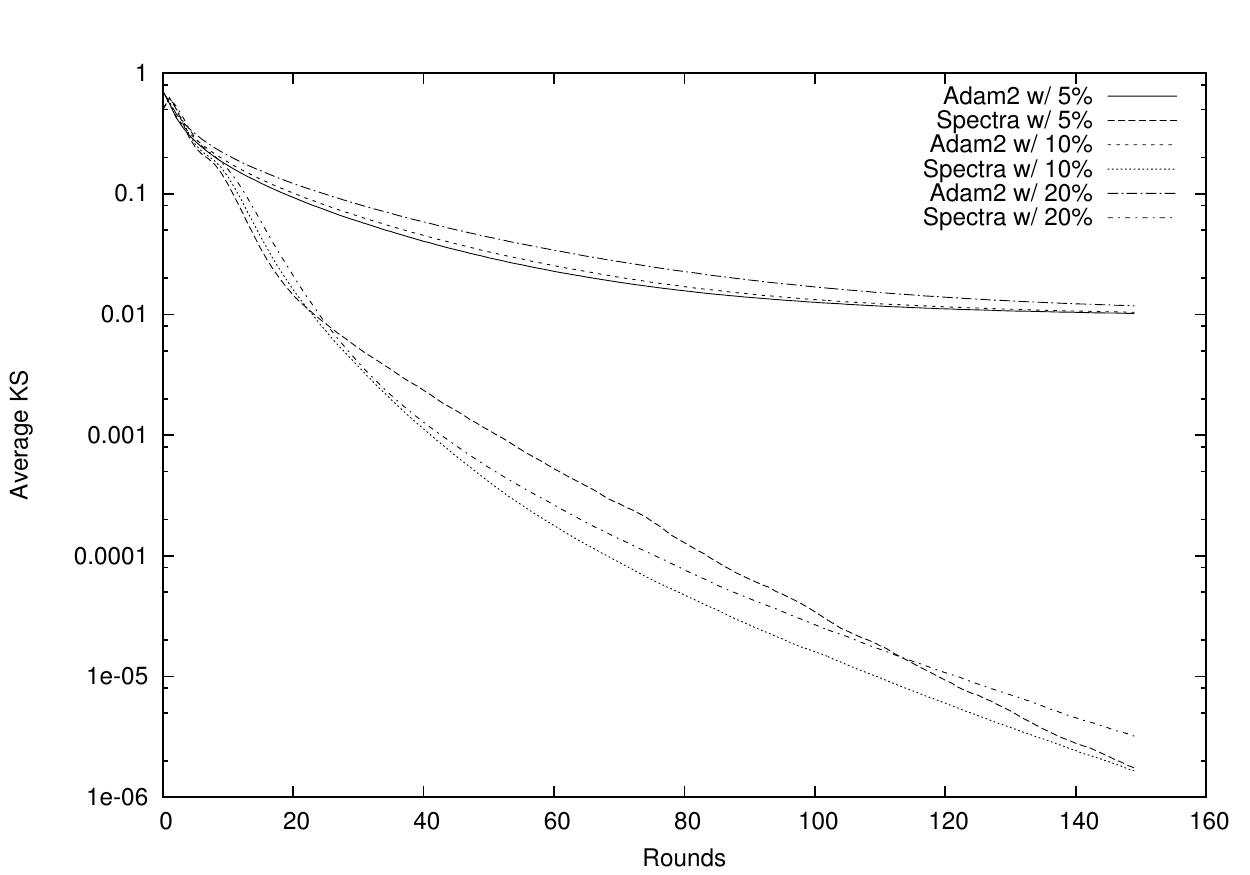}
\label{fig1:graf3}
}
\subfigure[Average KS error on a 1000 nodes random network running \textit{Spectra} with disturbance on initial values.]{
\includegraphics[width=.45\textwidth]{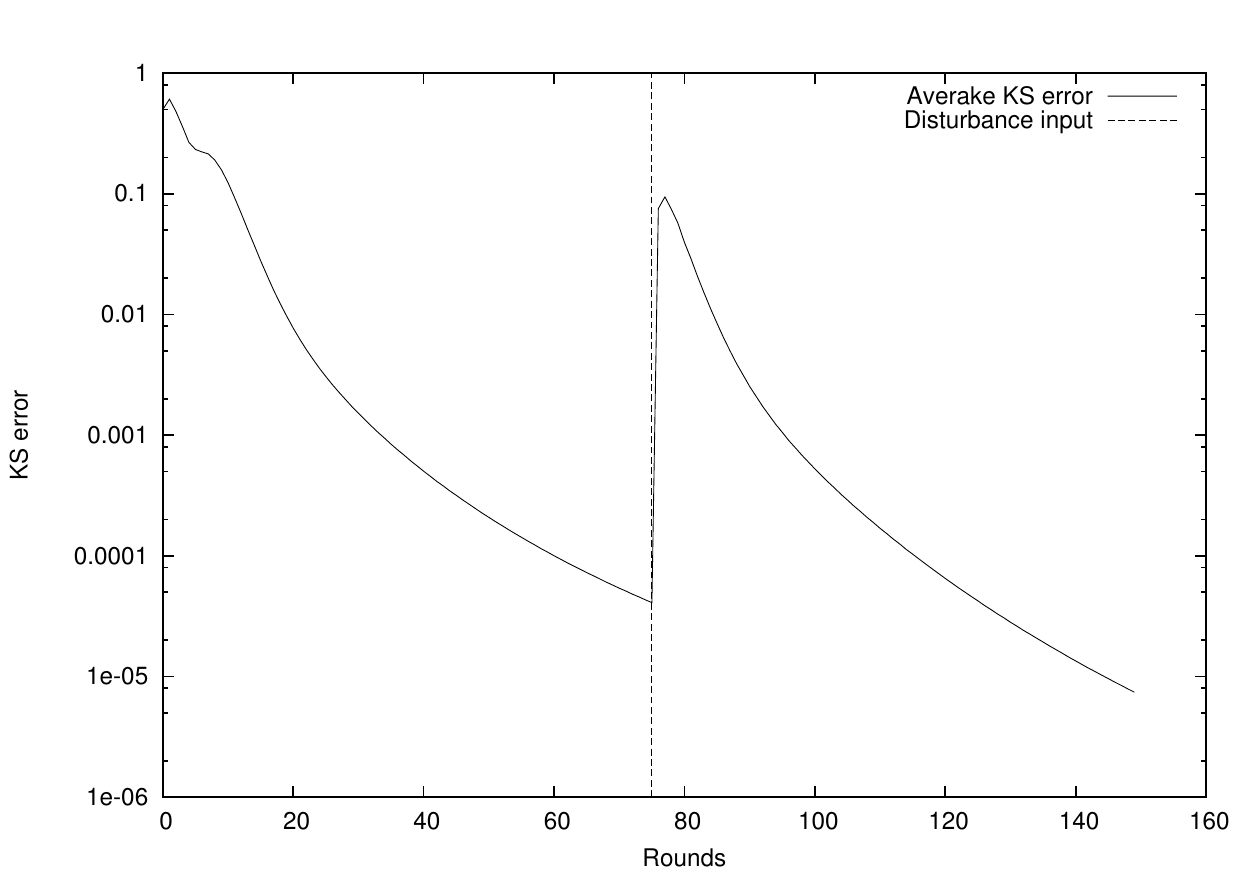}
\label{fig2:graf5}
}
\subfigure[Average KS error, maximum KS error and node count on a random network running \textit{Spectra} while subjected to churn.]{
\includegraphics[width=.45\textwidth]{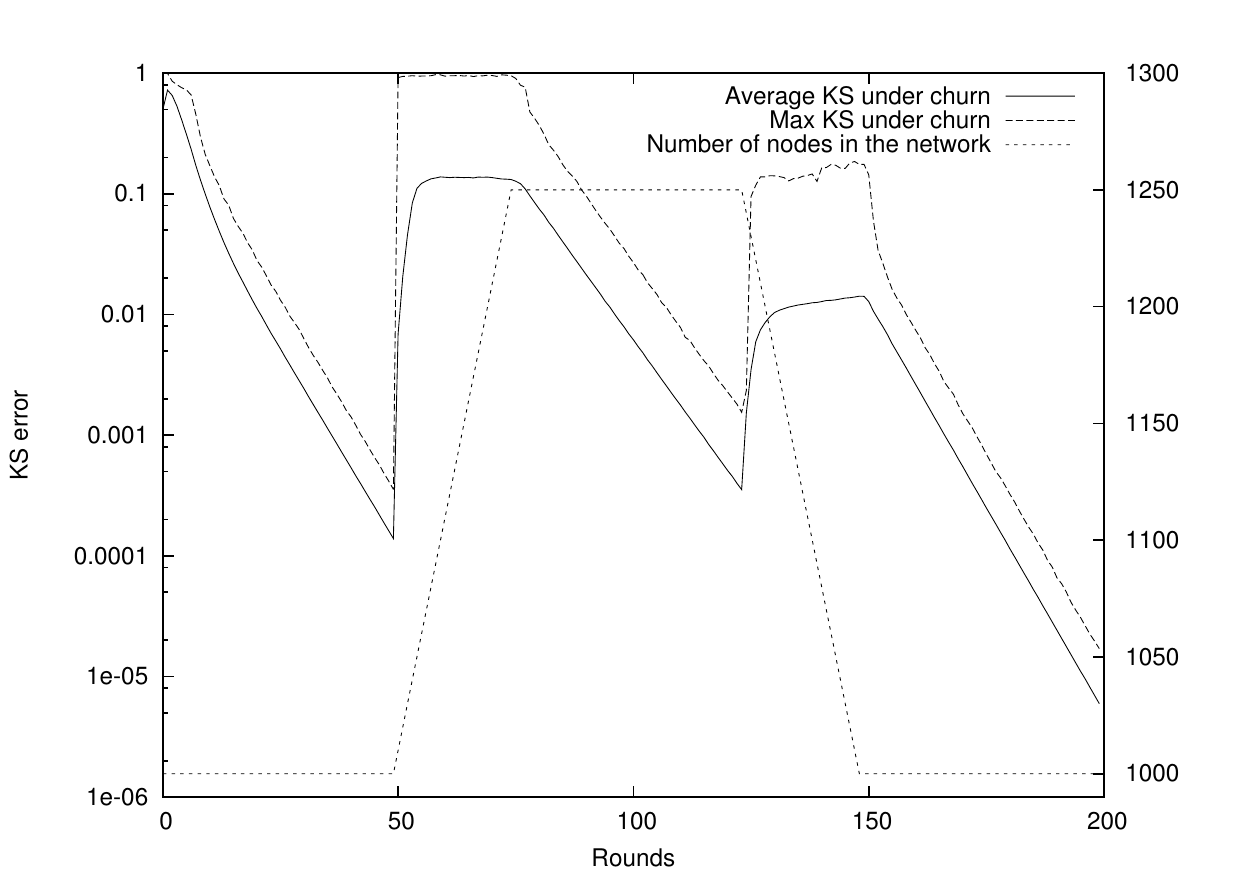}
\label{fig2:graf6}
}
\label{fig:subfigures}
\caption{Simulation results \subref{fig:adam2spectra}, \subref{fig1:graf3}, \subref{fig2:graf5} and \subref{fig2:graf6}}
\end{figure}

Figure~\ref{fig:adam2spectra} presents a graph with the average Kolmogorov-Smirnov distance to the real distribution on the nodes (as per Equation~\ref{ksmed}). It shows the performance of
both algorithms. One can observe that \textit{Adam2} converges asymptotically
to a non-zero value with a continuous offset error while \textit{Spectra} converges
indefinitely to zero. Also, the convergence speed is notoriously higher in
\textit{Spectra}, with orders of magnitude smaller error.

\subsection{Fault tolerance}

In this scenario we have simulated message loss rates of 5\%, 10\% and 20\% in each round. Results are presented in~\ref{fig1:graf3}.

Regarding \textit{Spectra}, we notice a slower overall convergence rate with 5\% message loss when compared to the other loss rates. This indicates that the algorithm is not only
resilient to message loss but also that with a message loss rate of up to
20\%, the convergence rate improves. This result is coherent with results
obtained in~\cite{Jesus:2010ke}.

This emergent behavior is in a way contradictory with the intuition that
convergence performance should degrade with message loss increase.
Simulation results suggest the opposite. This behavior may be understood if
we look at message loss as momentary changes in the network topology (a lost
message is equivalent to the extinction of an edge during a round).  This
effect is justified by the fact that the number of cycles in the network
topology tend to deteriorate the convergence performance in the underlying
averaging algorithm~\cite{Jesus:2010ke}.

We have also applied \textit{Adam2} to the same rates of message loss. Loss
of messages seem to introduce a systematic offset error.

\subsection{Dynamic adaptation to changes}

In order to evaluate the algorithm's behavior under dynamics in the sampled
value, we have conducted a simulation  that introduces  a disturbance on 20\% of the nodes chosen
uniformly random at round 75. The disturbance increased sampled values in 10\%, with
care not to change the minimum nor maximum of the network. The issues
concerning changes in minimum and maximum and the way it affects the
estimate will be addressed in future work.

The obtained results illustrate the adaptive nature of the proposed algorithm (see Figure~\ref{fig2:graf5}).
Without need to restart the protocol, the error in the estimate increases
at the moment of the disturbance and quickly converges to error rates
similar to pre-disturbance values. This behavior stems from the
averaging protocol used underneath and to the way each node preserves its
own sampled value and relative position. Iteratively, as rounds progress,
all nodes adjust their estimates taking into account the perceived value
changes and the consequent flow adjustments.

In order to test the resilience of \textit{Spectra} to churn, we have submitted
the algorithm to the departure and arrival of new nodes. In particular, it starts with a
network of 1000 nodes and at round 50, the number of nodes starts to
linearly increase (a 1\% increase per round), up to 1250 nodes at round 75. Then, after 50 rounds of
stability, nodes randomly leave the network at the same rate until it reaches again
1000 nodes. Node departure is equivalent to nodes crashing, as they leave silently 
without notifying any neighbors. In order to prevent network partitioning, the average node degree has been increased to 7 (i.e. $\approx ln(1000)$), following what has been done in  ~\cite{Jesus:2010ke}.
Data presented in Figure~\ref{fig2:graf6} depicts the average KS (as per Equation~\ref{ksmed}) and maximum KS error (as per Equation~\ref{ksmax})
for the whole procedure. It also depicts a profile with the number of nodes that constitute the network throughout the rounds. 
The arrival of nodes introduces new values to the distribution. The estimation has to be adjusted and thus the surge in the error levels. As soon as the node number stabilizes, the error levels decrease. Node departure introduces a similar effect.
These results show the algorithm's adaptability to
high churn rates and how quickly it converges to near zero error. The
estimates are computed uninterruptedly, without any need to restart the
algorithm \-- this property makes it highly adaptable and fault tolerant.

\section{Conclusions and Future Work}
	We have presented a distributed algorithm that computes the estimate of
cumulative distributed functions over a large scale network. The proposed
algorithm, \emph{Spectra}, overcomes the problems that previous approaches
exhibited. Our solution converges to the correct distribution, even when
facing high levels of message loss and churn in the network membership and
topology. It also allows dynamic adaptation to changes in the monitored
values (and their distribution), and avoids the need to re-start the
algorithm and loose progress. 

All the nodes have access to a high precision estimate of the CDF, and can
infer the associated distribution function. This data, being richer than
more simple statistics (e.g. average) allows a precise characterization of
the target network property and permits more accurate control decisions in
the presence of outliers and skewed distributions. 

As future work we intend to evolve the technique in order to allow for
variations in the minimum and maximum of the target property, since
currently we only adapt to growing maxima and decreasing minima. Another
improvement is in allowing an adaptive growth in the number of sampled
intervals, $k$, that is fixed at present. Finally we plan to address
strategies for consistent placement of the sample points, that will be no
longer uniform across the min-max range, as this will permit increased
sampling in areas where the changes in the property are more expressive, and
further increase precision.

\bibliographystyle{plain}
\bibliography{spectra_techreport_2012}

\end{document}